\documentclass[preprint,5p]{elsarticle}
\makeatletter
\def\ps@pprintTitle{%
 \let\@oddhead\@empty
 \let\@evenhead\@empty
 \def\@oddfoot{}%
 \let\@evenfoot\@oddfoot}
\makeatother

\usepackage{subfigure}
\usepackage{breakurl}
\usepackage{booktabs}
\usepackage[english]{babel}
\usepackage{multirow}
\usepackage{amsmath}
\usepackage[hang,flushmargin]{footmisc}
\usepackage{caption}
\usepackage{hyperref}

\makeatletter
\def\blfootnote{\xdef\@thefnmark{}\@footnotetext}
\makeatother

\begin{document}
	\begin{frontmatter}
		
	\title{Distribution System Monitoring for Smart Power Grids with Distributed Generation Using Artificial Neural Networks}
	\author[unik]{Jan-Hendrik Menke\corref{mycorrespondingauthor}}
	\ead{jan-hendrik.menke@uni-kassel.de}
	
	\author[unik]{Nils Bornhorst}
	
	\author[unik,iee]{Martin Braun}
	
	\cortext[mycorrespondingauthor]{Corresponding author}
	\address[unik]{Department of Energy Management and Power System Operation, University of Kassel, Wilhelmsh\"oher Allee 71-73, 34121 Kassel, Germany}
	\address[iee]{Fraunhofer Institute for Energy Economics and Energy System Technology, K\"onigstor 59, 34119 Kassel, Germany}

	\begin{abstract}
	The increasing number of distributed generators connected to distribution grids requires a reliable monitoring of distribution grids. Economic considerations prevent a full observation of distribution grids with direct measurements. First approaches using a limited number of measurements to monitor distribution grids exist, some of which use artificial neural networks (ANN). The current ANN-based approaches, however, are limited to static topologies, only estimate voltage magnitudes, do not work properly when confronted with a high amount of distributed generation and often yield inaccurate results. These strong limitations have prevented a true applicability of ANN for distribution system monitoring. 
	The objective of this paper is to overcome the limitations of existing approaches. We do that by presenting an ANN-based scheme, which advances the state-of-the-art in several ways: Our scheme can cope with a very low number of measurements, far less than is traditionally required by the state-of-the-art weighted least squares state estimation (WLS SE). It can estimate both voltage magnitudes and line loadings with high precision and includes different switching states as inputs. Our contribution consists of a method to generate useful training data by using a scenario generator and a number of hyperparameters that define the ANN architecture. Both can be used for different power grids even with a high amount of distributed generation.
	Simulations are performed with an elaborate evaluation approach on a real distribution grid and a CIGRE benchmark grid both with a high amount of distributed generation from photovoltaics and wind energy converters. They demonstrate that the proposed ANN scheme clearly outperforms state-of-the-art ANN schemes and WLS SE under normal operating conditions and different situations such as gross measurement errors when comparing voltage magnitude and line magnitude estimation errors.
	\end{abstract}
	

	\end{frontmatter}

		
	\section{Introduction}
	
	There is an increasing amount of DG installed on the medium and low voltage levels. Since PV and WEC power injections are volatile, unlike those of conventional power plants, the power flows of those grids become increasingly unpredictable regarding both the magnitude and the direction. To postpone grid reinforcements as long as possible, grids are often operated closer to their operational limits. The knowledge of the grid's state is imperative to prevent violation of operational limits. This creates a demand to monitor distribution grids in a reliable way. To monitor a grid, various measurement devices have to be installed. SE, which has been used in transmission grids for decades, can estimate current grid states by evaluating the measurements. It requires a large number of redundant measurements which are often not viable to gather in distribution grids. Both the installation of measurement devices and the required communication infrastructure are very expensive and may thus be uneconomic at lower voltage levels.
	
	\blfootnote{
		Abbreviations:\\
		ACC: accuracy class\\
		ANN: artificial neural network\\
		DG: distributed generator\\
		DSSE: distribution system state estimation\\
		MV: medium voltage\\
		PFC: power flow calculation\\
		PV: photovoltaic\\
		SD: standard deviation\\
		SE: state estimation\\
		SR: success rate\\
		STV: scenario tuple value\\
		WEC: wind energy converter\\
		WLS: weighted least squares
	}
	
	Many different grid monitoring methods have been developed over the last decades, some especially suited for distribution systems under the aforementioned conditions. The standard approach is the WLS SE \cite{abur2004power}, which was developed in the early 1970s. A redundant number of measurements is processed so that the weighted squared error between a calculated state and the measurements is minimized. Heuristic search methods can be used to solve the state estimation problem, as it is done by Mobash and El-Hawary in \cite{7726788}. If the required number of measurements is not available, pseudo measurements are often used to supplement the existing measurements. A different method is forecast-aided SE \cite{5233865}, which seeks to incorporate forecasts directly into the estimation process. In our previous work \cite{hmmpaper}, we have developed a method more suitable for DSSE.
	
	While pseudo measurements are sometimes required (e.g., for WLS SE), and can increase the accuracy of an estimation \cite{6176289}, their generation requires additional information, e.g., about weather conditions or detailed load profiles. This necessitates processing prior to the estimation itself. Ideally, a method forgoes the mentioned requirements while still yielding the desired accuracy.
	
	In this paper, we research a different approach which is not based on the classical state estimation. The approach utilizes ANN to build a model of the electrical grid through extensive training. After the model has been trained to sufficient accuracy (resulting from the user's individual requirements), grid voltages or line currents can be calculated almost instantly by multiplying the measurement vector with the ANN's trained weighting matrix.
	
	Artificial neural networks have been used in different power system applications (e.g., see overview by Hassan~et al. in \cite{HASSAN2013134}) and for power system monitoring. 
	Early approaches, such as \cite{Garcia-Lagos2000} from Garc{\'i}a-Lagos~et~al., consider ANN as the solver for the WLS SE optimization problem. In \cite{1460111}, Khoa~et~al. present two different methods for SE: Both a neural network and a parallel genetic algorithm estimate the state vector by minimizing the WLS fitness function. In \cite{1338474}, Singh~et~al. perform DSSE with ANN on the IEEE\,14 system and 2000 load cases which differ up to $\pm15\,\%$ in power.
	In a more recent approach, Mosbah~et~al. use a similar methodology for the same grid in their paper \cite{7379974} from 2015. 
	No information about measurements (types, numbers, positions, accuracies) used for training and validation are given. 
	Ivanov and Garvrila\c{s} perform the estimation of bus voltage magnitudes with an ANN in \cite{6420026}, although the estimation accuracy is relatively low with $> 2\,\%$ error in most cases. SE is performed via ANN on an under-determined, accurate set of measurements by Onwuachumba~et~al. in \cite{6520082}. In \cite{7051666}, they perform the same process on a grid with a dynamic topology. The error is considerably larger for an untrained topology. Barbeiro~et~al. train an autoencoder ANN to predict state values from measurements in \cite{Barbeiro2015108}.
	In \cite{6860997} and \cite{6947718}, Ferdowsi~et~al. examine the use of ANN in low voltage distribution feeders.
	The same authors use the approach on a MV grid in \cite{7066896}. The test grid contains different types of loads (residential, industrial) and three switches, but no DG. 
	In \cite{7066896}, they account for different switching configurations by using multiple ANN, requiring a high computational effort. Pertl~et~al. investigate monitoring of a low voltage benchmark grid with high DG penetration for two validation cases in \cite{7741758}. Most recently, Weisenstein~et~al. use ANN to monitor LV grids with smart meters in \cite{neis2018ann}. The large estimation errors make it infeasible for use in real operating conditions.
	
	The main limitations exhibited by state-of-the-art approaches using ANN can be summarized as follows:\\
	1. The training data used in existing ANN approaches is not chosen in a way to deliver accurate results for today's and future distribution grids with a high amount of DG, energy storage, electro-mobility, and so on.\\
	2. The estimation accuracy in state-of-the-art ANN approaches is still low.\\
	3. For each possible switching state, a separate ANN is required, yielding a high computational effort.\\
	4. Only voltage magnitudes are estimated. However, information about line current magnitudes is crucial for detecting line overloading.\\
	The objective of this paper is to overcome these strong limitations, thus enabling a true applicability of ANN for distribution system monitoring. Our ANN-based scheme consists of the following key contributions: The first key contribution is the controlled generation of training data. We propose a scenario generator generating scenarios for ANN training such that the ANN training is appropriate also for a high amount of DG. Finding usable hyperparameters for ANN is crucial to achieving high accuracy. As the second key contribution, we present a set of hyperparameters, which yield low errors in our tests and can be adapted to be used on other power grids with a different number of available measurements. The third contribution is the inclusion of switch statuses as inputs to the ANN, which are trained for normal operation cases. Unusual switching states, which can appear, e.g., in fault cases, are not accounted for. However, due to low training times for the ANN (several minutes), new ANN can be trained on the fly to include the previous and the new switching states if required. This has the disadvantage that no accuracy can be predicted in advance for the new switching state, but that would also be the case with other monitoring schemes. The forth, minor contribution is the introduction of an error correction method for voltage measurements. Line overloading is a serious danger in grids with high penetration of DG, therefore we also estimate the line current magnitudes in addition to bus voltage magnitudes.
	
\begin{figure}
		\centering
		\includegraphics[width=0.48\textwidth]{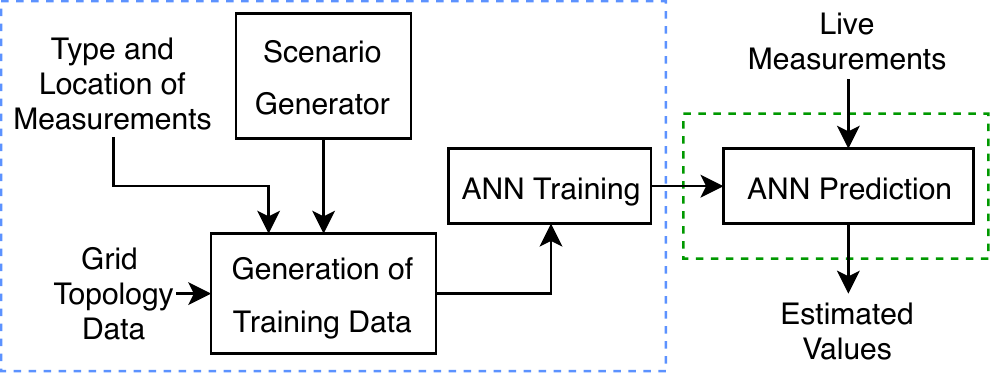}
		\caption{Flow chart of the scheme. The blue box on the left shows the preparation phase while the green box represents the operation phase in which live measurements are used for the trained ANN to predict the desired values.}
		\label{fig:mfc}
	\end{figure}

	The scope of this paper includes the introduction of the new ANN-based scheme, specifically the ANN architecture and the required training data, the test of the scheme using over 30 test cases with thousands of test scenarios on a benchmark grid, and the additional validation on a real distribution grid containing many PV generators and WEC. The general methodology of our presented scheme from training to operational use is detailed in Fig.~\ref{fig:mfc}.
	
	We shortly introduce the term artificial neural network in Section~\ref{ann-chapter}. In the following Section~\ref{sec:seWithAnn}, we detail the process of grid monitoring with ANN and introduce the scenario generator. The test grids are presented in Section \ref{sec:networks}. The evaluation with different grids is performed in Section~\ref{sec:evaluation}.

	\section{Artificial Neural Networks}
	\label{ann-chapter}
	
	An ANN is a computational system which consists of individual elements, called neurons. Neurons possess connections to other neurons, through which information can be exchanged. A perceptron, which is a commonly used model for a neuron, receives $n$ different inputs $x_1\ldots{}x_n$. The inputs are weighted by weights $w_i$, summed up and then passed through an activation function (e.g., the Rectified Linear Unit-function $y=f(x)=\mathrm{max}(0, w\cdot{}x)$). The perceptron's output $y$ is returned. A single perceptron is the simplest form of a neural network.
	Typically, an ANN consists of multiple parallel perceptrons, which form a layer, and one or several layers connected serially. The outputs of all neurons in the previous layer are passed to each neuron in the next layer. This configuration is called multi-layered perceptron. For an in-depth explanation of neurons and multi-layered perceptrons, see \cite{natureofcode}.
	
	ANN can adapt to a problem by training with an amount of data. Training takes place in multiple epochs. In each epoch, the training dataset is fed into the current grid configuration and an optimizer adjusts the weights to better fit the desired outputs. A separate validation dataset can be used to stop the training before overfitting occurs.
	
	\section{Grid Monitoring with ANN}
	\label{sec:seWithAnn}
	
	Electrical power grids can be modeled with different components and its states described by the power flow equations. With enough known inputs, these equations can be used to calculate the state variables of the system (complex bus voltages), from which all power flows and line currents can be derived. This process is called PFC and is regularly done in both simulations and control centers.
	
	Distribution grids on the low and medium voltage level are usually not equipped with enough measurements to directly perform a PFC or state estimation. The system of equations is under-determined. Contrary to WLS SE, ANN can be used as function approximators without having a fixed requirement for the minimum number of measurements.
	
	Therefore, through proper training, a neural network can be used to find a mapping from different measurements to the variables which should be monitored.
	
	\subsection{Monitoring under Real Operating Conditions}
	
	To be usable for distribution grids under real operating conditions, a monitoring method should function under different aspects: It has to function with a limited set of measurements and still deliver results within the desired accuracy. It also has to work despite occurring topology changes. The runtime must be below a certain time limit (e.g., 1 second). Additionally, the method should be robust against bad data (measurement loss ...).
	
	\subsection{ANN Architecture}
	
	A number of parameters have to be chosen in a specific way before grid variable approximation can be performed successfully with an ANN. These are mainly the number of ANN layers, the amount of neurons per ANN layer, the activation functions of the different layers, the weight initialization, the ANN connection structure, an adequate set of scenarios to approximate the internal function of the grid via training, and the optimization algorithm for adjusting the network's weights.
	
\begin{figure}
		\centering
		\includegraphics[width=0.4\textwidth]{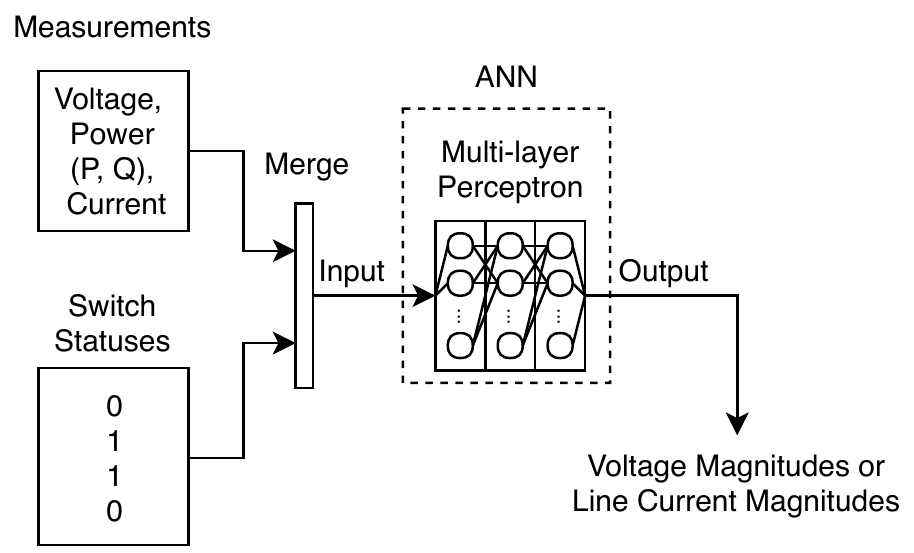}
		\caption{Data flow into and out of the ANN.}
		\label{fig:ann-flow}
	\end{figure}

	In this paper, we use a multi-layered perceptron architecture (cf. Section \ref{ann-chapter}), which is a recommended approach for supervised regression problems in \cite{Goodfellow-et-al-2016}. One ANN estimates the bus voltage magnitudes, a second network estimates the line current magnitudes. The data flow to and from the ANN are shown in Fig.~\ref{fig:ann-flow}. The input layer size is equal to the number of measurements plus the amount of switches. The output layer size is equal to the number of bus voltage magnitudes or the amount of line current magnitudes, respectively.
	The amount of neurons per hidden layer is set as
	\begin{equation}
	\label{eq:layersize}
	n_{\mathrm{hidden}}=[\frac{2}{3}\,n_{\mathrm{in}}+n_{\mathrm{out}}],
	\end{equation}
	where $n_{\mathrm{in}}$ is the size of the input layer and $n_{\mathrm{out}}$ the size of the output layer.
	This formula has been suggested as an initial layer size in \cite{heaton2008introduction}. Using this value as the starting point, the ideal layer sizes, which depend on the specific problem, are found by tuning the hyperparameters of the ANN model. This can be done using established methods, e.g., grid search or random search. All layers are fully connected with bias. Three hidden layers are used in the ANN, which results from an investigation in Subsection \ref{ssec:layeramounts}. The output layer has linear activation, the other layers use the ReLU function for activation. We have also investigated other activation functions (sigmoid, tanh). However, the above-mentioned combination yields the best results in our simulations. The initial weights of the neural network are chosen according to the Bottou proposition \cite{Thimm1995}: A uniform random sample is drawn from $[-2.38\,n_l^{-0.5}, 2.38\,n_l^{-0.5}]$, where $n_l$ is the size of the layer. Regularization methods like batch normalization or dropout were tested as well, but are not used in our model due to unsatisfactory results.
	
	The inclusion of switching states as inputs follows from an assessment: Networks can be operated in different switching configurations. If a grid contains $s$ switches, then, in theory, there are $2^s$ switching states possible. While many states are implausible (e.g., all switches open), the amount of valid states nevertheless grows significantly with the number of switches. Instead of retraining up to $2^s$ networks individually if, e.g., a line is modernized or a DG added, which is required in the approach of \cite{7066896}, only the single ANN has to be retrained in our approach.
	
	The training data consists of a number of training samples and features for each sample (total size of the training data: samples\,$\cdot{}$\,features). Each sample corresponds to a scenario. The features are an array of the used measurements and the switch positions. The training process of the ANN lasts for 100\,-1000 epochs, depending on the grid. A validation set, which is built from 25\,\% of the training data, can stop the training process early to prevent overfitting. The optimization algorithm \textit{Adam} is chosen to optimize the ANN's weights with a learning rate of 0.001 such that the mean squared error between the ANN's output and the reference output is minimized. Training data is generated by using scenarios from the scenario generator (Subsection~\ref{ssec:scenariogeneration}), performing a PFC to generate the full system states, and applying Gaussian noise individually to account for the tolerance of measurement devices. Additionally, the Gaussian noise acts as regularization for the ANN training, helping to prevent overfitting.
	
	\subsection{Scenario Generator}
	\label{ssec:scenariogeneration}
\begin{figure}
		\includegraphics[width=0.5\textwidth]{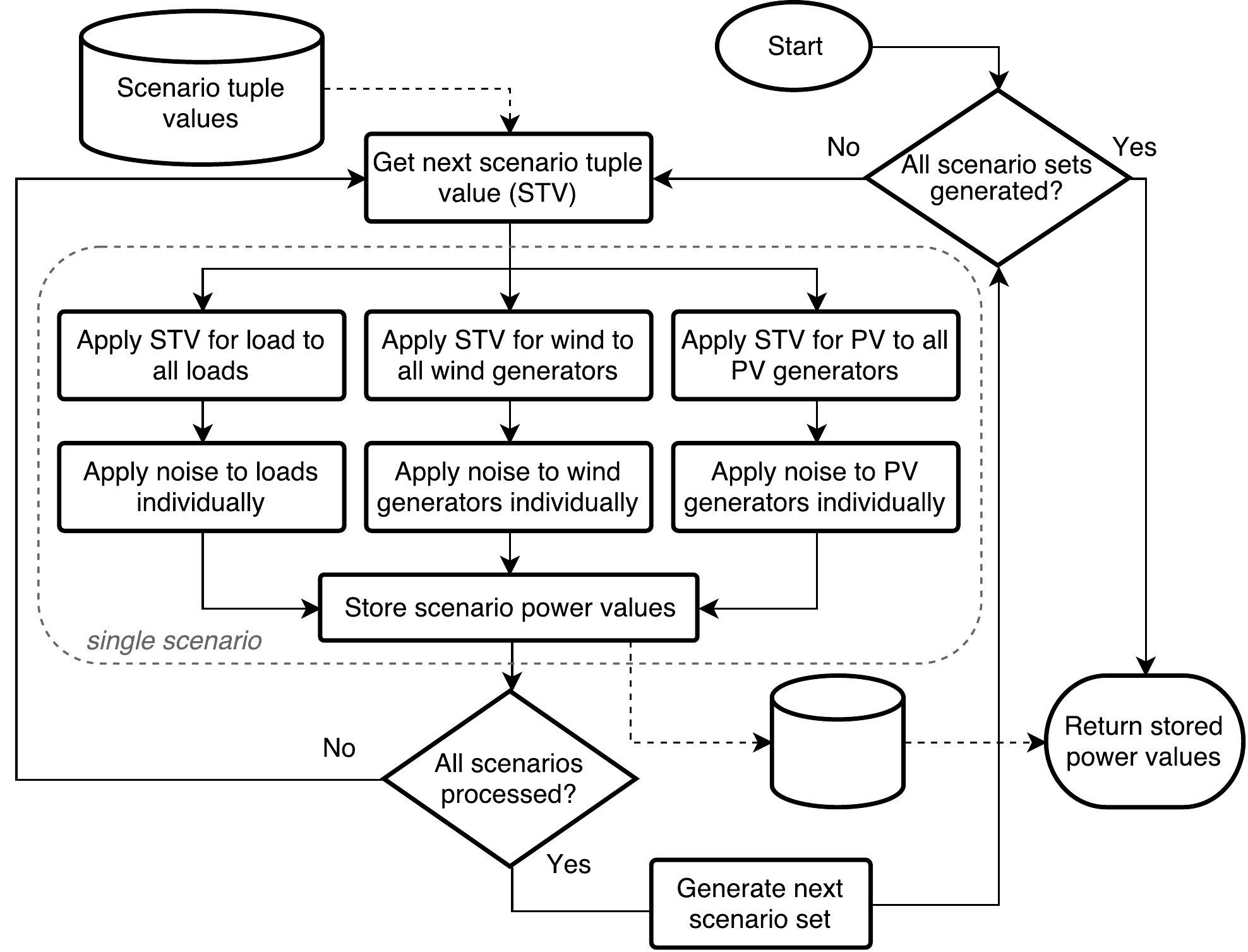}
		\caption{Process to create set of scenarios. A scenario consists of power values for all load and generators. (0.9, 0.7, 0.8) is an example for a STV.}
		\label{fig:power-value-gen}
	\end{figure}

	For the ANN to properly approximate the relationship between measurements and the output values, it is important to train with a large amount of diverse situations which can arise in the grid. In state-of-the-art ANN schemes, training is performed with only a selected number of scenarios, e.g., the extreme scenarios as in \cite{7741758}. In our simulations, we show that this approach yields an insufficient accuracy (see Subsection~\ref{ssec:sota-training}). A very important aspect of our successful simulations is not only the model but the scenario generator. Sampling and training from the whole range of possible scenarios enables the ANN to interpolate between the trained scenarios well and in turn estimate the relevant variables with high accuracy. Our scenario generator aims to cover as many situations as possible at a reasonable computational effort. To this end, in our tests, we consider three different parameters regarding the bus power injections: the load, PV generator power, and WEC power. These parameters are independent of each other. Load power is scaled from 10\,\% to 100\,\%, WEC power from 0\,\% to 100\,\%, and PV generator power from 0\,\% to 90\,\% , since a base value of 100\,\% PV power is unrealistic according to \cite{denavns} and load profiles generally do not dip below 10\,\%. 
	A scenario consists of a tuple of the scaled values for these three types.
	The scenarios are used as base values for the test cases. Assuming a significant correlation between all loads, between all PV generators, and between all WEC, respectively, each scenario consists of only a single load scaling value, a single PV generator scaling value, and a single WEC scaling value. Each load is then scaled with the load scaling value, each PV generator with the PV generator scaling value, and each WEC with the WEC scaling value. To account for the variability among individual units of the same type, all units are then further scaled individually with Gaussian noise. Gaussian noise with a SD of $25\,\%$ ($10\,\%$ noise SD for loads) is applied. The noise values for DG have been chosen by interpreting various sources, e.g., \cite{HOFF20101782}, and adding a generous margin to capture differences between DG units of the same type. The $\cos \varphi$ is automatically set for all scenarios to match the $\cos \varphi$ of the given nominal loads ($0.97$ in our scenarios). Fig.~\ref{fig:power-value-gen} details the process of converting the scenarios into power values for all loads and generators.
	
	One request to the scenario generator, i.e., one repetition in Fig.~\ref{fig:power-value-gen}, yields a set of $10\cdot{}11\cdot{}10=1100$ scenarios.
	The next request yields a different set of scenarios due to the Gaussian noise applied to each load, WEC, and PV generator individually.
	
	Due to the variability of parameters between 0\,\% and 100\,\%, different scenarios exhibit large differences in the summed bus power injections. Hence, a sufficient variability in the scenarios for distribution grids with a high amount of DG is achieved with the scenario generator for training and validation.

	\subsection{Extending the Method}
	Beyond the scope of this paper, possible extensions of the method increase its applicability in practice: For unbalanced grids, the proposed approach can easily be extended by increasing the number of input variables (measurements) and output variables (grid variables) of the ANN by a factor of $n$ for $n$ phases. The training and monitoring process is identical.
		
	Voltage angles can be estimated in addition to voltage magnitudes, e.g., to enable full state estimation and the execution of subsequent methods like optimal power flow. This can be achieved by estimating the voltage angles using an additional ANN and will be part of our future work.
	
	\section{Electrical Test Networks}
	\label{sec:networks}
	
	The estimation accuracy of the ANN monitoring is evaluated with two test grids: 
	
	The first grid is a slightly modified version of the CIGRE MV benchmark grid found in \cite{1709447}, which is available open-source with the power system analysis tool \textit{pandapower} \cite{pandapower}. It contains 15 buses, 8 PV generators, and 1 WEC; the voltage level is $20\,kV$. The 2 batteries, 2 fuel cells and 2 combined heat power plant devices are neglected. The nominal power of all remaining DG has been doubled as to create more problematic grid situations and compensate for the neglected types of DG. This grid is tested with a set of 4 different switching configurations. Table~\ref{tab:cigre-switches} shows the used switching configurations.
	
	\begin{table}[b]
		\centering
		\caption{Switching configurations for the CIGRE test grid.\\1: closed; 0: open.}
		\label{tab:cigre-switches}
		{\footnotesize
			\begin{tabular}{@{}c|cccccc@{}}
				\toprule
				\multirow{2}{*}{Configuration Number} & \multicolumn{6}{c}{Switches (connect bus\,A-bus\,B)} \\
				                     & 14-8 & 6-7 & 4-11 & 8-9 & 3-4 & 3-8 \\ \midrule
				0                    & 0    & 0   & 0    & 1   & 1   & 1   \\
				1                    & 0    & 1   & 0    & 1   & 0   & 1   \\
				2                    & 1    & 0   & 1    & 0   & 1   & 0   \\
				3                    & 1    & 0   & 0    & 1   & 1   & 0   \\ \bottomrule
			\end{tabular}
		}
	\end{table}
	
\begin{figure}
		\centering
		\includegraphics[width=.5\textwidth]{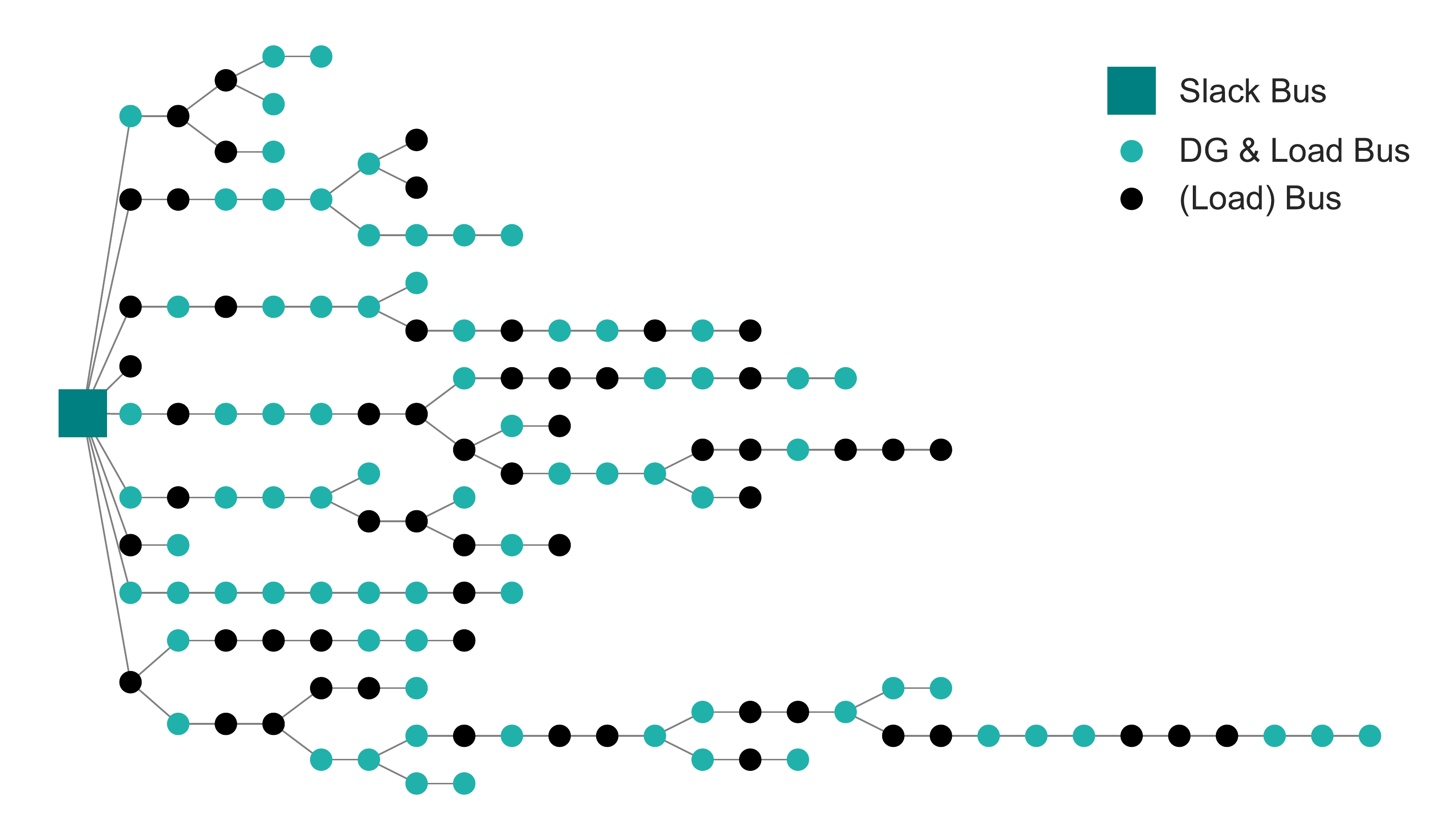}
		\caption{The real test grid with high DG penetration.}
		\label{fig:net1-switch-confs}
	\end{figure}

	The second grid is a real distribution grid on the $20\,kV$ level in Germany. Its model contains 135 buses, 134 lines, 82 loads, 75 PV generators and 2 wind farms. We have selected it for our simulations as the nominal DG power is almost 3 times the nominal load. The grid is operated in a radial fashion. The measurement positions and types are transferred from the real grid: 19 buses contain P, Q and V measurements. Additionally, the line flows of the feeders going out from the substation are measured (P, Q, I). This translates to $m=84$ real measurements for the grid with $n=137$ buses (a measurement redundancy of $\frac{m}{2n-1}=0.31$). Five different, plausible switching configurations have been selected for the test. A visual representation of the grid with one switching configuration is shown in Fig.~\ref{fig:net1-switch-confs}.
	
	Since the measurements which are used as input data are generated by the scenario generator and PFC, ACC, which describe the noise level of measurements, have to be chosen. As per IEC~61869, the ACC~0.5 is chosen for voltage measurements. The corresponding SD is $\frac{0.5}{3}\,\approx{}0.167\%$, which means that $99.7\,\%$ of measurements have a maximum error of $0.5\,\%$ ($3\,\sigma$ confidence interval). ACC~1 with a maximum error of $1.5\,\%$ is chosen for current measurements (SD of $0.5\,\%$). Power measurements, as a sum of voltage and current accuracy, are set with a SD of $\frac{2}{3}\,\%$.
	A wide range of simulations is performed to assess the viability of monitoring via ANN. 	
	
	\section{Simulations}
	\label{sec:evaluation}
	
	\subsection{Implementation \& Hardware}
	
	The implementation of the ANN is done in the programming language Python, using the \textit{PyTorch} \cite{pytorch} package.
	The required PFC are performed using pandapower \cite{pandapower}. WLS SE is also performed by using the state estimator from pandapower. The maximum number of iterations for WLS SE is set to 10. An additional stopping criterion is the minimum state update of $10^{-6}$ between two iterations. The existing measurements are supplemented by substitute values (pseudo measurements) for bus power injections to provide the required number of measurements to the SE. 
	Substitute values are created with a simple method by summing the generated power inside the grid and dividing it among buses according to each load's installed power. Advanced methods for the generation of pseudo measurements, e.g., using Gaussian mixture models \cite{singh2010distribution, pieri_pseudomeasurements}, typically require additional information (customer profile time series, historical data, algorithms upstream of the WLS SE \cite{6176289} ...). We seek to compare the methods using the same data by transferring a very basic, but commonly available, strategy from an industry control center software. To create substitute values, unmeasured WEC and PV generators are given the same share of power injection as the measured DG of the same type, e.g., the measured relative power $P_\mathrm{solar_{rel}}=P_\mathrm{solar}/P_\mathrm{solar_{max}}$ is transfered to all other generators of the same type. The total DG power in the grid, both measured and estimated, is summed as $P_\mathrm{dg}$. The active load $P_u$ of unmeasured buses $u$ is set as
	\begin{equation}P_u=(-P_\mathrm{slack}-P_m-P_\mathrm{dg})\cfrac{P_{\mathrm{max}_u}}{\sum_u P_{\mathrm{max}_u}}\label{eq:loadsub2},\end{equation} where $P_\mathrm{slack}$ denotes the active power injection of the slack bus, $P_m$ is the sum of measured loads, and $P_{\mathrm{max}_u}$ the maximum installed load of bus $u$. Therefore, each load is assigned a share of the power injected into the grid, scaled by the load's installed power. The reactive power is set according to a defined $\cos \varphi$, 0.97 in our scenarios.
	The substitute values are added with a SD of $30\,\%$ of the value to account for their imprecise nature. The substitute values are only used by the WLS SE, so that P and Q measurements are available for every bus and the grid is observable.
	All results are produced with an Intel i7-4702MQ CPU (2,2\,GHz), 16 GB of RAM (800\,MHz), SSD storage, Python 3.6 on Ubuntu Linux without using GPU acceleration.

	\subsection{Evaluation Approach}
	\label{sec:evaluationapproach}
	
 	The new monitoring method should be characterized under different situations like normal operation, bad data, etc.. As a reference, the results are compared with the industry standard WLS SE method. Both monitoring algorithms are tested with the same data and are evaluated by the same criteria. A comparison to state-of-the-art ANN schemes is not shown since they are not applicable as demonstrated in Subsections \ref{ssec:sota-training}\,/\,\ref{ssec:sota-arch}. 
 	
	The aim is to compare the methods in multiple situations to determine their impact on the approximation accuracy. There are several different conditions to check for: In normal operation conditions, both the amount and the types of measurements impact the monitoring quality. These parameters are varied in different test cases. The behavior of the monitoring method under influence of bad data is also important and tested with different types of errors (gross measurement errors, measurement outages, and topology errors).
	
	Each test case is assigned an ID, e.g., M0, to identify it in the results. ``Voltage magnitude measurements'' are shortened to ``$V$ measurement'' in the following subsections. ``$I$ measurements'' refer to ``line current magnitude measurements''. $S_{\mathrm{bus}}$ stands for complex bus power injection measurements while $S_{\mathrm{line}}$ denotes complex line flow measurements.
		
	The variations of measurement configurations are shown exemplarily in Table~\ref{tab:scenariomeasconfigurations} for the first test grid as the example case. The characteristics of the test cases which contain bad data and topology errors are shown in Table~\ref{tab:scenariocharacteristics}.
	
	\begin{table}
		\centering
		\caption{Measurement configurations for all test cases (by ID).}
		\label{tab:scenariomeasconfigurations}
		{\footnotesize
			\begin{tabular}{@{}lllll@{}}
				\toprule
				ID & $V$   & $S_{\mathrm{bus}}$   & $S_{\mathrm{line}}$ & $I$ \\
				\midrule
				\multicolumn{5}{l}{\textit{Variation of Measurements}}                   \\
				M0       & 0           &             & 1-2, 12-13           &            \\
				M1       & 0, 6        &             & 1-2, 12-13           &            \\
				M2       & 0, 6, 10    &             & 1-2, 12-13           &            \\
				M3       & 0, 6, 8, 10 &             & 1-2, 12-13           &            \\
				M4       & 0, 6, 8, 10 & 4, 7        & 1-2, 12-13           &            \\
				M5       & 0, 6, 8, 10 & 4, 7, 9, 11 & 1-2, 12-13           &            \\
				M6       & 0, 6, 8, 10 & 3 - 11      & 1-2, 12-13           &            \\
				M7       & 0, 6, 8, 10 & 3 - 11      & 1-2, 8-9, 12-13      &            \\
				M8       & all         & all         & 1-2, 4-5, 8-9,       &            \\
				         &             &             & 3-8, 6-7             &            \\
				M9       & 0, 7        & 0, 7        & 1-2, 12-13           &            \\
				\midrule
				\multicolumn{5}{l}{\textit{Variation of Measurement Types}}              \\
				A0       & 0           &             &                      & 1-2, 12-13 \\
				A1       & 0, 6, 8, 10 & 4, 7        &                      & 1-2, 12-13 \\
				A2       & 0           &             & 1-2, 12-13           & 1-2, 12-13 \\
				A3       & 0, 6, 8, 10 & 4, 7        & 1-2, 12-13           & 1-2, 12-13 \\
				\midrule
				\multicolumn{5}{l}{\textit{Other Test Cases}}                            \\
				F0 - F7  & 0, 6, 8, 10 & 4, 7        & 1-2, 12-13           &            \\
				P0 - P3  & 0, 6, 8, 10 & 4, 7        & 1-2, 12-13           &            \\
				T0 - T5  & 0, 6, 8, 10 & 4, 7        & 1-2, 12-13           &  			 \\
				\bottomrule         
			\end{tabular}
		}
	\end{table}
	
	\begin{table}[]
		\centering
		\caption{Description of the individual test cases (by ID) regarding bad data and topology errors.}
		\label{tab:scenariocharacteristics}
		{\footnotesize
			\begin{tabular}{@{}ll@{}}
				\toprule
				ID & Characteristic                                                   \\
				\midrule
				\multicolumn{2}{l}{\textit{Bad Data / Gross Measurement Errors}}      \\
				F0 & V measurement at bus 0 is 0 p.u.                                 \\
				F1 & V measurement at bus 8 is 0 p.u.                                 \\
				F2 & PQ measurement at line 1-2 is 0 W / Var                          \\
				F3 & PQ measurement at line 12-13 is 0 W / Var                        \\
				F4 & V measurement at bus 8 is 150\,\% of the real value              \\
				F5 & PQ measurement at bus 4 is 150\,\% of the real value             \\
				F6 & Measurement accuracy known to estimator is incorrect:            \\
				& Accuracy of PQ measurement at bus 4 is 6\,\% instead of 2\,\%       \\
				& Accuracy of V measurement at bus 8 is 3\,\% instead of 0.5\,\%      \\
				F7 & V measurement outage at bus 8                                    \\
				& Use of constant substitute value: $1.0\,\mathrm{p.u.}$              \\
				P0 & PQ power at bus 4 is only 70\,\% of the measurement value        \\
				P1 & PQ power at bus 4 is 130\,\% of the measurement value            \\
				P2 & PQ power at buses 5, 9, 10 are 130\% of the measurement value    \\
				P3 & Both P0 and P2                                                   \\
				\midrule
				\multicolumn{2}{l}{\textit{Topology Errors}}                          \\
				T0 & R, X of line 7-8 only 90\,\% of actual value                     \\
				T1 & R, X of lines 3-8, 7-8, 8-9 only 90\,\% of actual value          \\
				T2 & Switch status of switch 2 inverted                               \\
				T3 & Switch status of switches 2 and 3 inverted                       \\
				T4 & R, X of all lines inaccurate in uniform range: {[}90\,\%, 110\,\%{]} \\
				T5 & Both T2 and T4                                                   \\
				\bottomrule                      
			\end{tabular}
		}
	\end{table}
	
	In total, 32 test cases are used.
	Three different sets of scenarios (3300 scenarios) from the scenario generator in Subsection \ref{ssec:scenariogeneration} are used for testing, i.e., a repetition limit of three is chosen in Fig.~\ref{fig:power-value-gen}. Although our analysis of bad data and measurement errors is not exhaustive, we test for many different fault conditions which can appear in operation. An exhaustive investigation of fault conditions is beyond the scope of this paper.\\
	Instead of focusing on specific numbers to compare the results of the different methods, we define approximations as \emph{successful} or \emph{unsuccessful}. An approximation is defined as successful if all errors are below a predefined limit. In this case, a grid operator would know the scope of uncertainty when interpreting the monitoring method's result. Two evaluation criteria, C1 and C2, are defined in Table~\ref{tab:simcriteria}. The SR is the ratio of successful estimations to the number of total estimations.
	
	\begin{table}[b]
		\centering
		\caption{Evaluation criteria for the test cases. The error is the difference between the value which is estimated using the noisy measurements and the true value, which is the result of a power flow calculation with no noise applied.}
		\label{tab:simcriteria}
		{\footnotesize
			\begin{tabular}{ll}
				\toprule
				Criterion C1: & If both the bus voltage magnitude errors \\
				& are less than $1\,\%$ and line loading errors\\
				& are less than $10\,\%$, the scenario is defined\\
				& as successful.\\ 
				Criterion C2: & Strict version of C1, with limits of $0.5\,\%$ \\
				& for voltage magnitude errors and $5\,\%$ for \\
				& line loading errors. \\
				 \bottomrule                                                             
			\end{tabular}
		}
	\end{table}
	
	\subsection{Evaluation for the CIGRE Network}
	\label{sec:se-comparison}
	
	\subsubsection{State of the Art Training Method}
	\label{ssec:sota-training}
	
	In \cite{7741758}, ANN are trained using 5 scenarios for the tested grid. The usability of this training method is evaluated with the test case M4, the CIGRE test grid and all of our test scenarios.
	
	The overall SR using only these training scenarios is 0.0\,\% and 0.0\,\% (C1/C2). We conclude that, for our test grid and our test scenarios, such a small number of training scenarios is not suitable to train an ANN in such a way that it estimates voltage magnitudes or line currents with sufficient accuracy as defined in Table~\ref{tab:simcriteria}.
	
	\subsubsection{State of the Art ANN Architecture}
	\label{ssec:sota-arch}
	In \cite{7066896}, Ferdowsi~et~al. present a monitoring method for voltage magnitudes based on ANN which operates on a MV grid with 14 buses (we use the CIGRE benchmark grid with 15 buses on the MV level). A single hidden layer with two neurons is used. The size of the input and output layers are given by the measurements and the number of buses, respectively. The sigmoid activation function and the Levenberg-Marquardt optimizer are used. We use the training and test data generated for our test case M4 and train the aforementioned architecture in the MATLAB Neural Network Toolbox, as is done in \cite{7066896} and similarly in \cite{7741758}. The SR for the state-of-the-art architecture are 78.48\,\% for C1 and 38.57\,\% for C2 for voltage. This is considerably lower than the of 100\,\%\,/\,98.5\,\% (C1\,/\,C2) achieved with our architecture.

	\subsubsection{Determining the Amount of Hidden Layers, Training Epochs, and Training Data}
	\label{ssec:layeramounts}
	
\begin{figure}
		\includegraphics[width=0.5\textwidth]{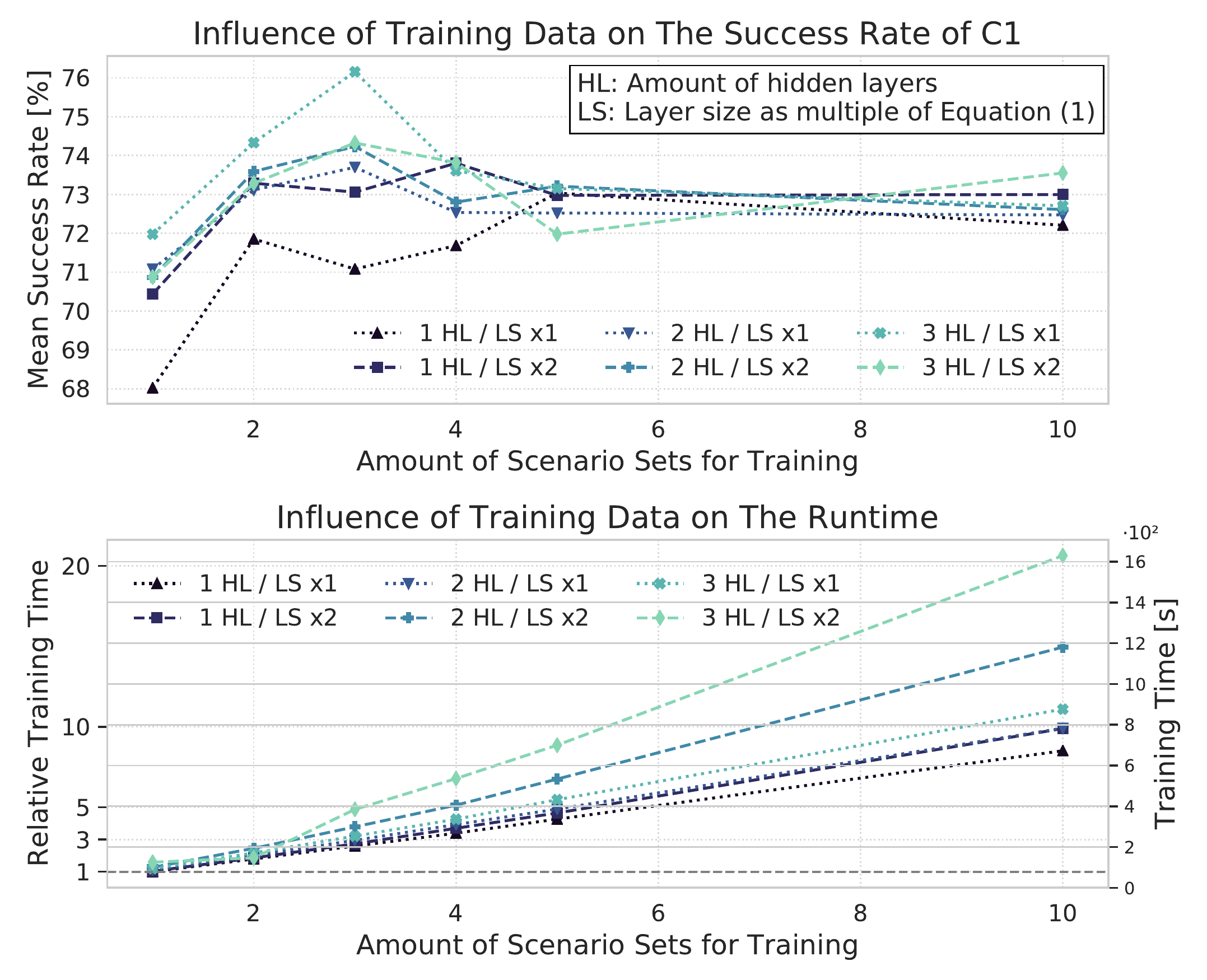}
		\caption{Top: The mean SR of C1 of all test cases defined in Table~\ref{tab:scenariomeasconfigurations} depending on the ANN architecture. Bottom: The relative training time of different ANN architectures as a mean training time of the test cases from Table~\ref{tab:scenariomeasconfigurations}, compared to the shortest training time. The absolute values are shown on the right y-axis in the bottom subplot.}
		\label{fig:ann-evaluation}
	\end{figure}

	The ideal number of hidden layers, training epochs and the amount of training data are determined using various combinations and the SR is recorded for each combination. Values between 100 and 10000 training epochs are simulated with various architectures. 150 training epochs were found to be sufficient for high accuracy. In Fig.~\ref{fig:ann-evaluation}, we evaluate the impact of hidden layers and the layer size on the average SR and training time. For simplicity, we have chosen integer multiples of the layer size of Equation~(\ref{eq:layersize}). We use the 32 test cases from Table~\ref{tab:scenariomeasconfigurations}, a total of $32\cdot{}3\,300=105\,600$ test scenarios, to calculate the mean SR of a single data point in Fig.~\ref{fig:ann-evaluation}. The resulting best combination w.r.t. SR, three hidden layers and three sets of training scenarios, is selected in the further simulations. This combination offers a reasonable trade-off between training time and SR.
	
	\subsubsection{Simulation Results}
	
	\begin{figure*}
		\includegraphics[width=\textwidth]{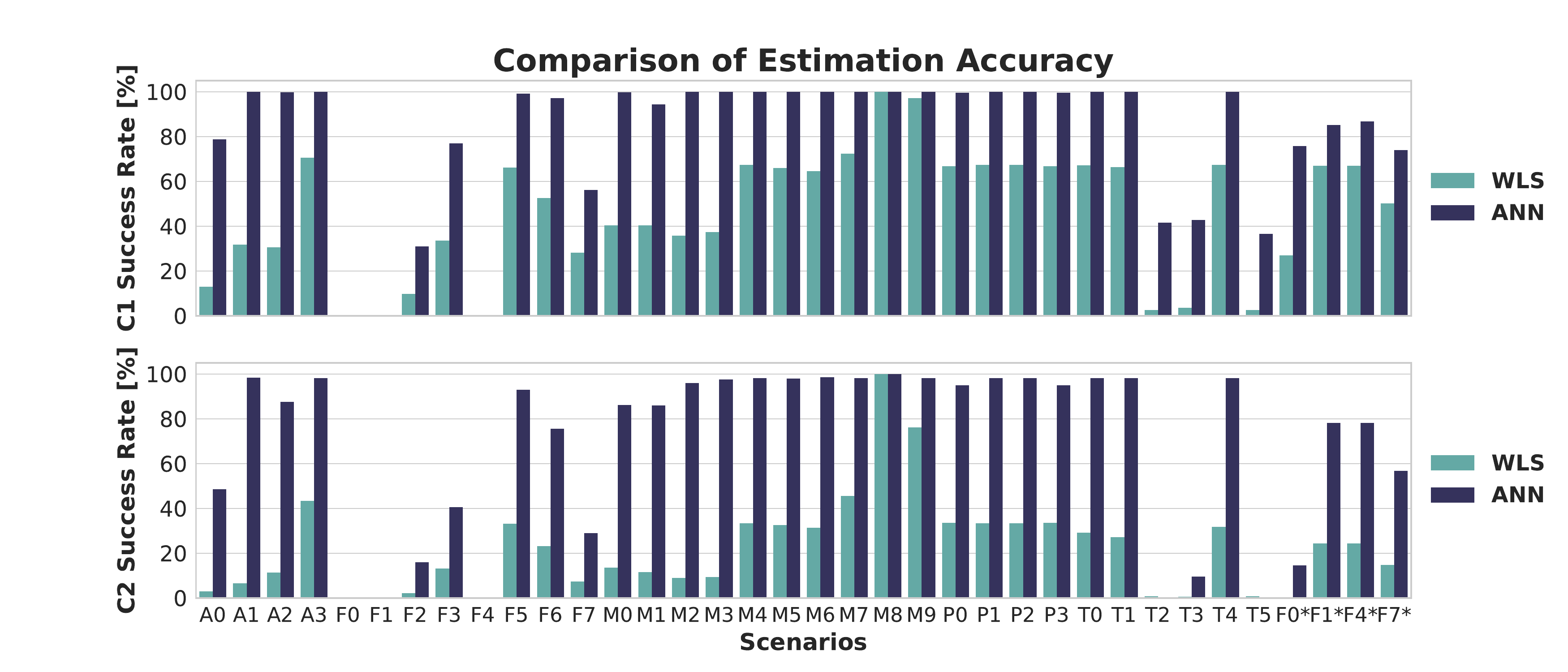}
		\caption{Results of the test cases of the CIGRE MV benchmark grid. The upper graph shows the successful approximation rate for the less strict comparison criterion C1, the lower graph shows the same result for the strict criterion C2.}
		\label{fig:cigre-results}
	\end{figure*}

\begin{figure}
		\centering
		\includegraphics[width=0.35\textwidth]{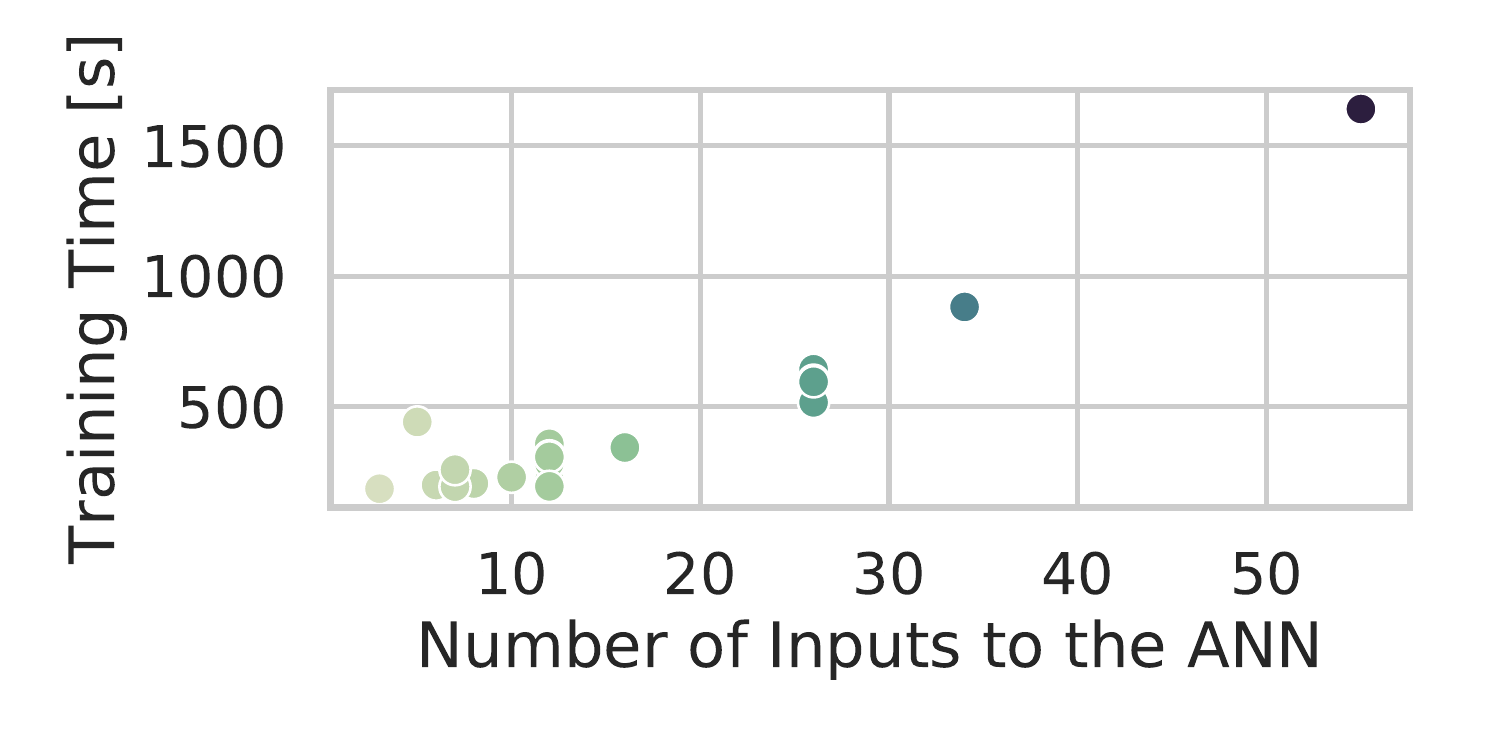}
		\caption{Training times for both ANN (estimating voltage magnitudes and line current magnitudes, respectively) combined per test case of Table~\ref{tab:scenariomeasconfigurations}. The test cases are identified by the number of ANN inputs. The results are produced with the computer described in the label of Fig.~\ref{fig:ann-evaluation}.}
		\label{fig:ann_training_times}
	\end{figure}

	The training times of the ANN pairs trained sequentially for the test cases of Table~\ref{tab:scenariomeasconfigurations} are shown in Fig.~\ref{fig:ann_training_times}. One ANN is used to estimate 15 bus voltage magnitudes, the other ANN estimates 15 line loading values.
	
	\begin{figure}
		\centering
		\includegraphics[width=0.5\textwidth]{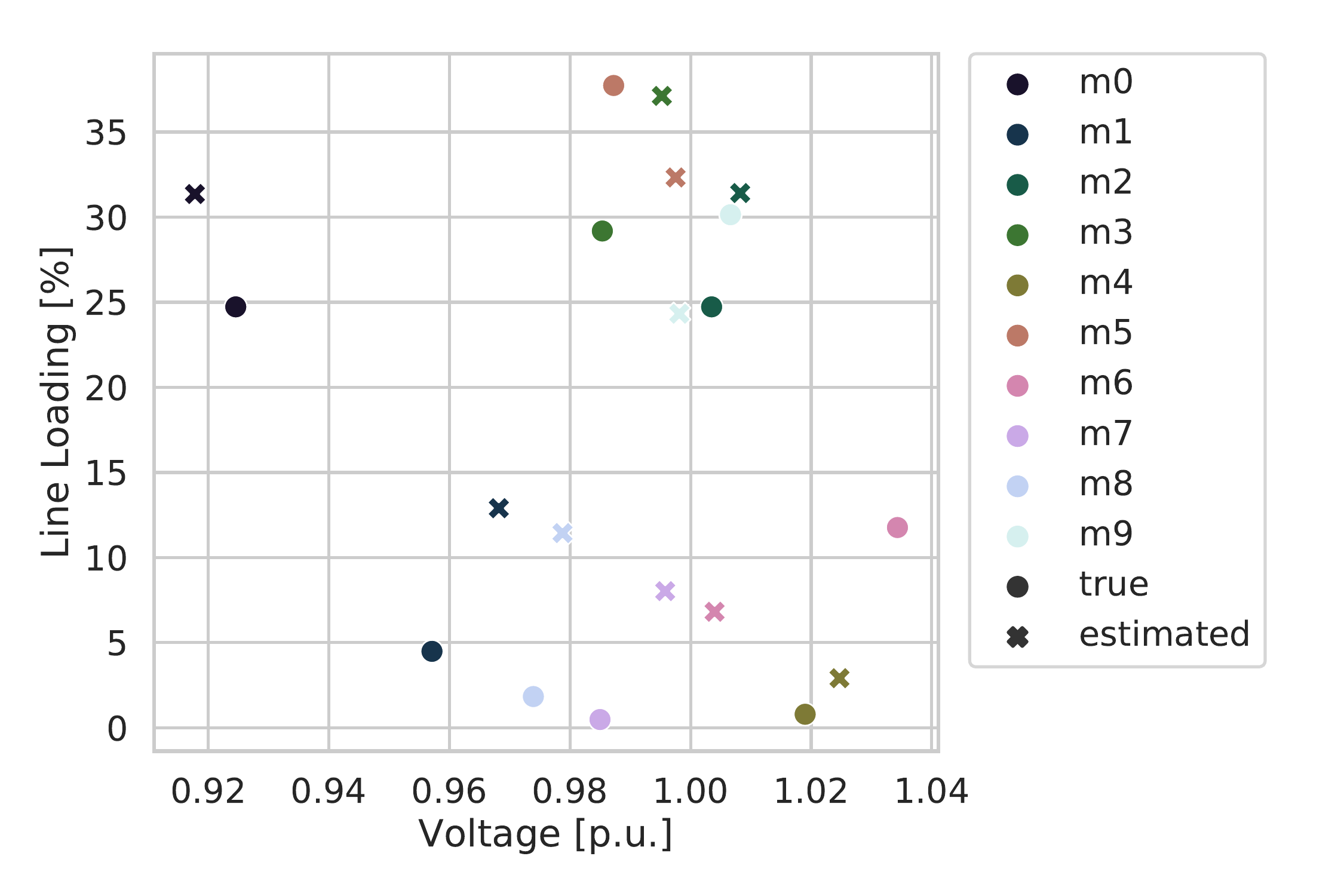}
		\caption{Maximum errors for the test cases M0 to M9. True values are plotted with circles; estimated values with crosses.}
		\label{fig:min_max_errors_m_cases}
	\end{figure}

	The SR for all test cases are shown in Fig.~\ref{fig:cigre-results}. The upper graph shows C1, the lower graph the C2 criterion. Firstly, we look at C1: For test cases regarding the variation of measurement, WLS SE has a SR of $35-70\,\%$. WLS SE only exhibits a relatively high SR if measurements are correct and there is a certain number of measurements available (M4~-~M7). Disadvantages are gross measurement errors (F0~-~F4) and topology errors (T2, T3, T5). Gross errors of bus power injection measurements in P0~-~P3 only minorly impact the SR. WLS exhibits a SR less than $73\,\%$ for all test cases except for M8/M9.
	
	The ANN method exhibits a high SR for most test cases and is the only method to reach a $100\,\%$ SR. It shares the disadvantages regarding measurement outages and incorrect switching positions with the other method. Without measurement or topology errors, only A0 and M1 produce a SR of less than $99.8\,\%$ for C1.
	
	The results for C2 show the same distribution of SR for the test cases, but an overall lower SR for both methods. For WLS SE, the gross measurement errors of bus power injections in P0~-~P3 have a higher impact on the accuracy of these test cases.
	
	Maximum estimation errors of the test cases M0 to M9 are shown in Fig.~\ref{fig:min_max_errors_m_cases}. While minimum errors almost exactly match the true values (error less than 0.1\,\%) and are therefore not plotted, the maximum errors show differences both regarding line loadings and voltage magnitude.
	
	The test case M8 is the result of tests to find a measurement configuration at which WLS SE reaches 100\,\% SR with the least amount of measurements. Measurements are added consecutively, first at the buses. After all buses are equipped with P, Q and V measurements, lines are equipped with P, Q measurements. As can be seen in Table~\ref{tab:scenariomeasconfigurations}, all buses are equipped with P, Q and V measurements. Additionally, five lines spaced evenly in the grid are equipped with P and Q line flow measurements. While the accuracy is met for voltage magnitudes with a significantly lower amount of measurements, the accuracy for line loadings requires almost full direct observation of the grid. This demonstrates that a significantly higher number of real measurements or a more sophisticated method to generate pseudo measurements is required for the WLS SE approach to increase its performance compared to the proposed ANN method.

\begin{figure}
	\includegraphics[width=0.5\textwidth]{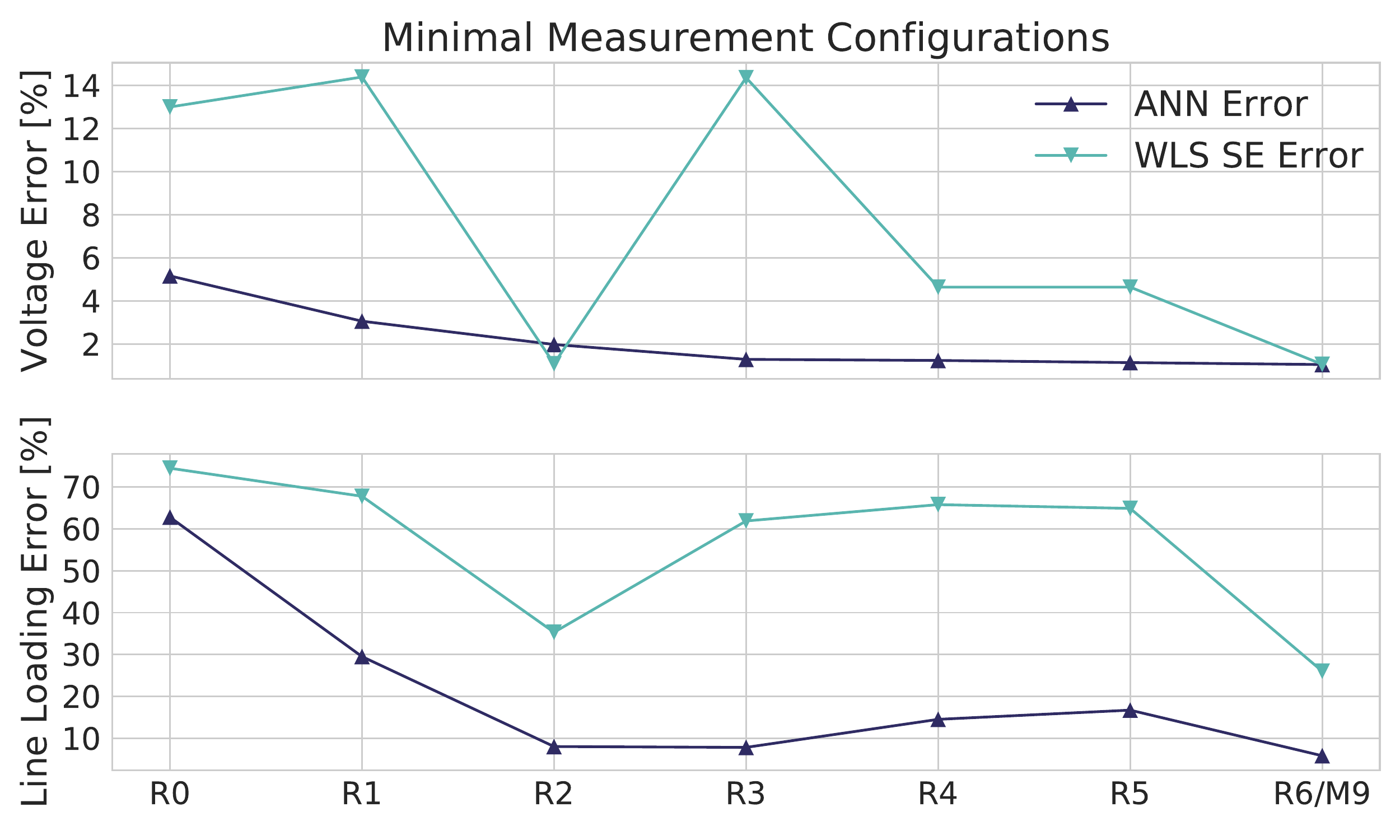}
	\caption{Voltage magnitude and line loading errors for the test cases R0-R6 to determine the required amounts of measurements to reach sufficiently low errors.}
	\label{fig:min-meas}
	\end{figure}

	Tests to find a minimal measurement configuration, which exhibits an acceptable voltage error of $\approx{}1\,\%$ and a line loading error of $\approx{}10\,\%$ in the worst cases, start with measurements only at the substation (bus~0). From a practical point of view, at the MV level, the substation is the bus where measurements exist most likely. Additional measurements are added either at the substation feeder ends (lines 1-2, 12-13), at the bus with the largest DG (bus~7) and at the end of a feeder (bus~11). All measurement configurations for the tests on reduced measurements are shown in Table~\ref{tab:minimalmeasurements}. Fig.~\ref{fig:min-meas} shows the results of the additional test cases for both the WLS SE and ANN monitoring method. The best test case, R6, is also shown as M9 in Fig.~\ref{fig:cigre-results}.
	
	\begin{table}
		\centering
		\caption{Minimal measurement configurations.}
		\label{tab:minimalmeasurements}
		{\footnotesize
			\begin{tabular}{@{}lllll@{}}
				\toprule
				ID & $V$   & $S_{\mathrm{bus}}$   & $S_{\mathrm{line}}$ & $I$ \\
				\midrule
				R0       & 0           & 0           &              &            \\
				R1       & 0           & 0           &              & 1-2, 12-13 \\
				R2       & 0           & 0           & 1-2, 12-13   &            \\
				R3       & 0           & 0, 7        &              & 1-2, 12-13 \\
				R4       & 0, 7, 11    & 0, 7, 11    &              &            \\
				R5       & 0, 7, 11    & 0           &              &            \\
				R6\,/\,M9& 0, 7        & 0, 7        & 1-2, 12-13   &            \\
				\bottomrule         
			\end{tabular}
		}
	\end{table}

	\subsubsection{Validation with Load Profiles and Battery Systems}
	
	The previous simulations were performed with scenario sets generated according to Subsection~\ref{ssec:scenariogeneration}. Due to the individual scaling of loads, PV generators, and WEC in the scenarios, the scenarios from the scenario generator exhibit a large variability. However, since the training and the evaluation scenarios are generated with the same scenario generator, it is interesting to investigate the performance of the proposed method with test scenarios which are not created by the scenario generator, e.g., with more realistic time series. Therefore, a further test with loads scaled according to German BDEW's H0 standard load profile and DG scaled according to the EV0 photovoltaic profile is performed. The wind power base values are chosen completely randomly in each time step. 864 scenarios (96 15-minute time steps for Winter, Summer, Transition days, each available for Weekday, Saturday, Sunday) are derived from the profiles, applied to each of the 4 switching configurations, and the test case M4 is repeated. The SR are $100\,\%$ (C1) and $99.97\,\%$ (C2) demonstrating that more realistic time series are even easier to handle for the proposed ANN method. The simulation results presented in the following paragraph further substantiate these findings.

	The CIGRE publication \cite{cigre_ext} presents time series for the distributed energy resources found in the MV benchmark along with two battery systems and loads that are distinguished by their profile (either residential or commercial). This increased complexity is incorporated in the following simulation: The two battery systems with a combined power of 800\,kW are added and either the residential or commercial profile are assigned as defined in \cite{cigre_ext}. The ANN is trained with scenarios from the scenario generator, using five axes (WEC, PV, battery, commercial load, and residential load) instead of the three axes used previously (WEC, PV, and load). Due to a large number of resulting training scenarios, the step size between 0\,\% and 100\,\% \text{(-100\,\% to 100\,\% for the battery axis)} is increased from 10\,\% to 20\,\% which results in considerably less training scenarios compared to 10\,\% steps. Moreover, only the reference switching state is considered.

	The trained ANN is tested with the 24-hour time series in 1-minute intervals from \cite{cigre_ext}. The battery is both acting as a load and as a generator at different times. The resulting SR are 100\,\% for C1 and C2. The max. voltage error is 0.25\,\% and the max. line loading error 1.8\,\%.
	All in all, the results show that different load profiles can be accommodated by incorporating the knowledge of the profile for each load into the training process.
	
	\subsubsection{Error Correction for Voltage Measurements}
	\label{ssec:verrcorr}
	
	In Fig.~\ref{fig:cigre-results} it can be seen that voltage measurement outages or constant replacement values for missing measurements can affect the SR heavily (F0, F1, F4, F7) while gross errors of bus power injections do not significantly affect the SR of the ANN method (P0~-~P3).
	
	To identify erroneous measurements, tests like the largest normalized residual test can be used in addition to basic plausibility checks. To deal with erroneous measurements, either the ANN can be retrained without the faulty measurement as an input or error correction methods can try to correct the measurement.
	
	Our approach to correct voltage measurement outliers, which is applicable independently of the monitoring method, sorts all available voltage measurements in ascending order. 
	The SD is computed for the set. Since gross errors will be found at the bottom or top of the sorted list, either measurement is removed from the set and the new SD is calculated. If the new SD is less than $50\,\%$ of the previous SD, the measurement is defined as an outlier and replaced by a substitute value. The substitute value is the mean of the measurement set without the outlier. This process is repeated until the SD does not fall below 50\,\% when removing a measurement.  This method is only applicable to voltage measurements since the values are close to 1.0\,p.u. in normal operation while power values vary significantly.
	
	The test cases F0, F1, F4, and F7 are repeated with the error correction method active. All test cases show increased SR, e.g., $86.73\,\%$ and $78.26\,\%$ for C1 and C2 instead of $0.0\,\%$ for F1*. The new SR are plotted in Fig.~\ref{fig:cigre-results} with a * next to the name. All results indicate that this simple error correction is effective in case of gross voltage measurement errors.

	\subsubsection{Recommendations for Measurement Infrastructure}
	From the results in the previous subsections, it is clear that monitoring of distribution grids can be performed reliably by using the ANN monitoring method. For our test grid with three types of energy resources (load, PV, wind), 6 measurements for 15 buses can be sufficient for reliable results as test case M3 shows. In general, with other test grids, gathering P, Q, and V measurement at approximately 15\,\% of buses additionally to measuring the substation feeder ends, has proven to be a good starting point for further optimization of measurement infrastructure. The measured buses should be those with a high rated power. Further requirements exist:
	It is crucial, not only for ANN but also for both monitoring methods, to correctly identify the switch statuses of the grid. Therefore, a topology processor should be used prior to the grid parameter approximation.
	Additionally, if measurement outages appear, they either have to be filtered from the measurement set beforehand, requiring retraining of the ANN, or error corrected if possible, e.g., by using the voltage error correction method presented in Subsection~\ref{ssec:verrcorr} if applicable.

	\subsection{Evaluation for the Real Test Network}
	
	After asserting the usability of the new ANN scheme with a benchmark grid and many test cases, the performance on the real test grid is investigated. The neural network weights were trained with the 2200 scenarios (2 sets of scenarios as per Subsection~\ref{ssec:scenariogeneration}), multiplied by 5 switching configurations. The accuracy of estimation is verified with 1100 different scenarios (5500 data vectors including the switching configurations). The voltage ANN was trained for 200 epochs, the line current ANN for 1000 epochs.
	
	Even though the grid topology is changed by the different switching configurations and the power demand of the grids varies between high load demand and high DG generation, the voltage magnitude estimation errors are below $0.43\,\%$ at all times ($-0.012\pm 0.05\,\%$). The line current magnitudes are estimated with an error margin of less than $9.8\,\%$ ($-0.1\pm 1.25$\,A).
	
\begin{figure}[]
		\includegraphics[width=0.5\textwidth]{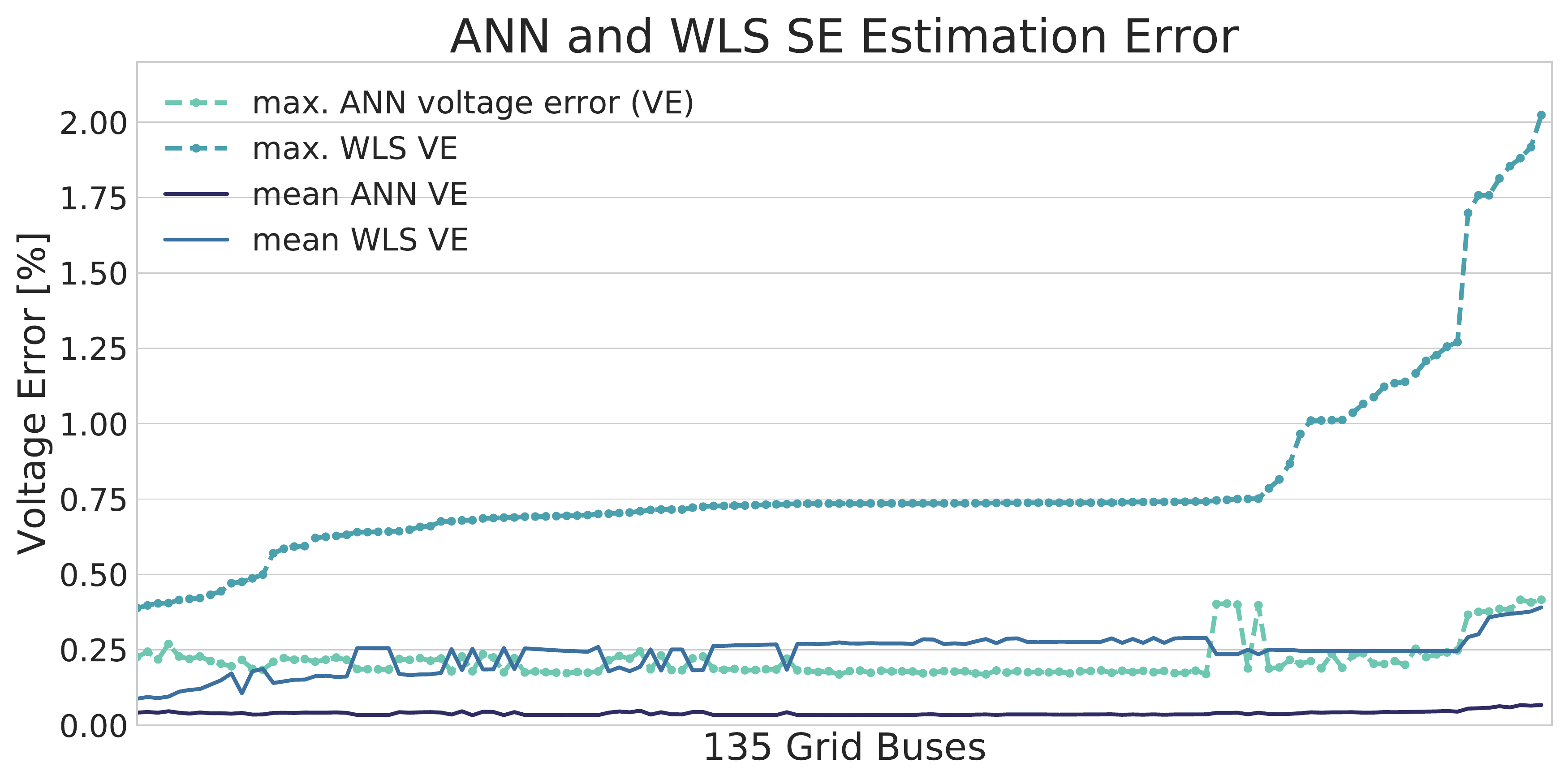}
		\includegraphics[width=0.5\textwidth]{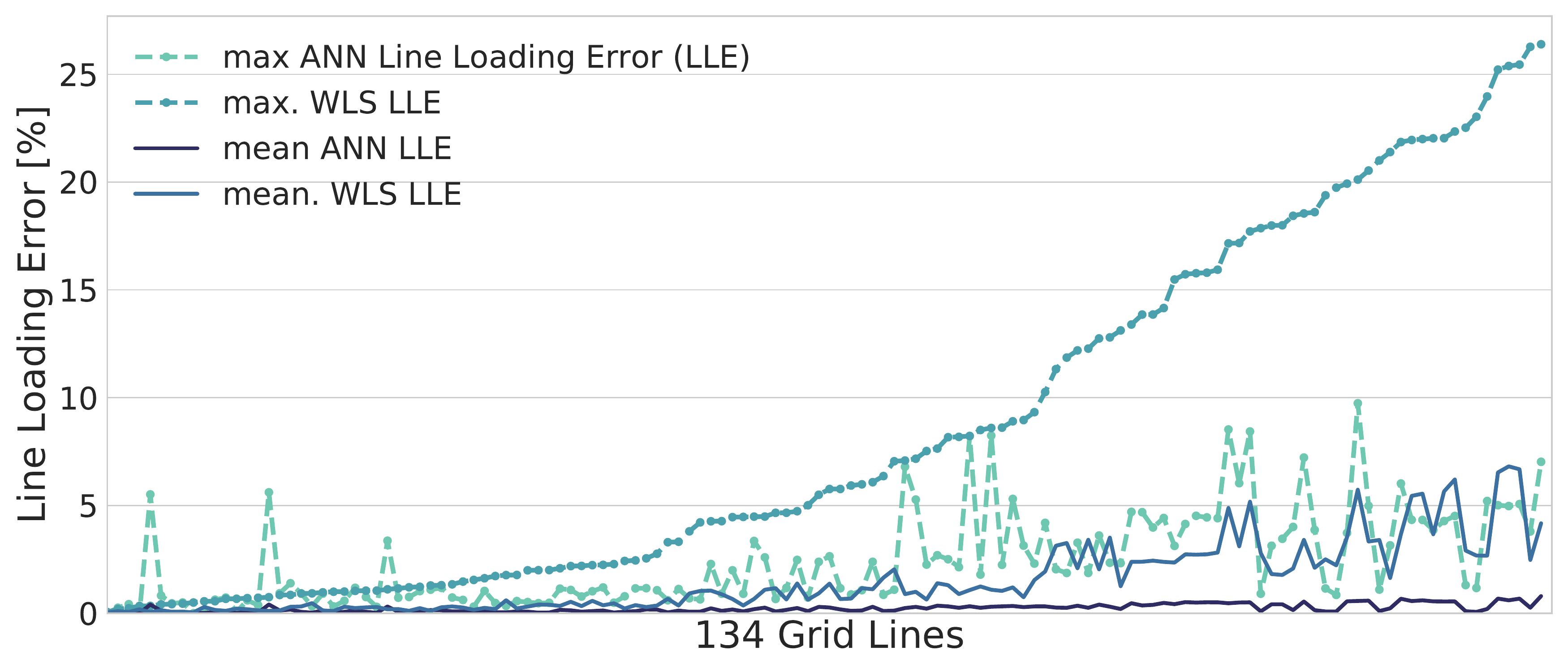}
		\caption{Comparison of mean and maximum voltage and line loading errors for both the ANN method and the WLS SE for all scenarios per bus/line. The bus and line errors are sorted by the maximum WLS SE error in ascending order.}
		\label{fig:schweigern-comp-err}
	\end{figure}

	The same simulations are also performed with the WLS SE as a reference. For the same measurements and test scenarios, the maximum voltage error of WLS SE is 2.02\,\% ($-0.146\pm 0.263\,\%$). The maximum error regarding line loading is 26.5\,\% (line current errors: $2.1\pm 9.76$\,A). The mean and maximum errors per individual bus and line are shown in Fig.~\ref{fig:schweigern-comp-err} for both methods.
	
	\subsection{Discussion of Results and Limitations}
	Based on the results of the previous subsections, the new ANN-based scheme presented in this paper overcomes many of the practical limitations of previous monitoring methods for distribution grids with a high number of DG and a low number of measurements. The estimation errors on both the benchmark grid and the real distribution grid are often low enough to fulfill the evaluation criteria C1/C2 and may thus be adequate for further use by grid operators.
	
	There are, however, limitations to the method: By adding more types of energy resources to a grid or monitoring unbalanced grids, the amount of training data increases significantly with the number of new types of energy resources or the number of additional phases in an unbalanced grid, which results in increased training times. Dynamic topologies, e.g., many switching states or transformers with taps, can also lead to a large number of training scenarios. The estimation of voltage angles does not yet work robustly. As a consequence, a full state estimation rather than a grid monitoring is not yet possible with the proposed method. A full state estimation is a requirement, e.g., for optimal power flow schemes.

	\section{Conclusion and Outlook}
	In this paper, we have proposed an ANN-based scheme to reliably perform distribution system monitoring with a limited number of measurements. The proposed approach overcomes limitations of existing ANN approaches which are not applicable to grids with high penetration of DG as clearly shown by our simulations. To appropriately train the ANN in the context of high penetration of DG and to test the reliability of the monitoring, a scenario generator which captures all possible grid states is presented. Results are both generated for a benchmark grid and a real distribution grid with a high amount of DG. WLS SE and ANN monitoring are compared under a wide range of situations. In all test cases, the ANN monitoring method outperforms the WLS SE using the same input data and the presented method for generating pseudo measurements. Although we concentrate on balanced grids in this paper, this is not a limitation. Either the process is performed in parallel for each phase for imbalanced grids or measurement data for all three phases is fed into a single ANN.
	
	Accurate monitoring of imbalanced grids as well as achieving full state estimation should be part of our future work. More complex power grids with a variety of customer load profiles and distributed energy resources also necessitate an approach to reduce the ANN training data. Considering practical use, the estimation accuracy should be increased in case of bad data, either by modifying the presented scheme or by selecting an appropriate method to identify and correct errors beforehand. The evaluation approach can be turned into an automatic general test process for monitoring methods. Parallel training of ANN can shorten the overall training time considerably.

	
	\section*{Acknowledgment}
	This work was partly supported by the German Federal Ministry of Economic Affairs and Energy and the Projekttr\"ager J\"ulich GmbH within the framework of the project OpSimEval (grant number 0325782B). The authors are solely responsible for the content of this publication. The authors kindly thank \textit{Netze BW GmbH} for their permission to use the topology data of a real distribution grid in our simulations.
	
	
	\bibliography{Literature}{}

\end{document}